\documentclass[5p,sort&compress]{elsarticle}
\usepackage{graphicx}

\usepackage{caption}
\usepackage{subcaption}

\usepackage[font=small,labelfont=bf]{caption}    

\usepackage{color} 
\usepackage{soul}

\usepackage{amssymb}
\usepackage{amsthm}
\usepackage{amsmath}

\usepackage{units}
\usepackage{bpchem}
\usepackage{bbding}

\biboptions{comma,sort&compress}

\journal{Scripta Materialia}

\newcommand{\be}{\begin{equation}}
\newcommand{\ee}{\end{equation}}

\begin{document}
\begin{frontmatter}

\title{Nanoporous Copper-Nickel -- Macroscopic bodies of a strong and deformable nanoporous base metal by dealloying}

\author[iww]{Lukas~L\"uhrs\corref{cor}}

\author[iww,hzg]{J\"org~Weissm\"uller}

\address[iww]{ Institute of Materials Physics and Technology, Hamburg University
of Technology, Hamburg, Germany}

\address[hzg]{Institute of Materials Research, Materials Mechanics, Helmholtz-Zentrum
Geesthacht, Geesthacht, Germany}

\cortext[cor]{Corresponding author. E-mail: lukas.luehrs@tuhh.de}
\begin{abstract}
Uniform macroscopic samples of nanoporous metal with high deformability
have so far been limited to precious metals such as Au, Pd and Pt. Here we propose nanoporous Copper-Nickel (npCN) as a nanoporous base metal that can be made with mm dimensions and exhibits significant deformability. NpCN forms a uniform bicontinous network structure with feature sizes that can be controlled from 13 to 40\,nm by thermal annealing. Continuous compression tests confirm ductile deformation behavior accompanied with a high strength compared to macroporous Cu- and Ni-foams with similar solid fraction.
\end{abstract}

\end{frontmatter}

\section{Introduction} \label{Introduction}

In the field of tailor-made nanomaterials, nanoporous metals made by dealloying promise significant functionalization due to their very large surface area and high structural integrity. Potential applications arise as actuators \cite{Jin2010,Detsi2012,Detsi2013}, sensors \cite{Stenner2016,Detsi2012}, catalysts \cite{Wittstock2010,Jin2011a}, microfluidic pumps \cite{Xue2014}, bioanalytical systems \cite{Seker2018} and structural materials with tunable mechanical properties \cite{Jin2011,Ye2013,Mameka2014,Sun2015,Luhrs2017}.
Three properties are generically important for the materials performance in these fields: Affordability is an obvious requirement. Furthermore, deformability is required for avoiding premature failure upon exposure to stress concentrations or to designed mechanical load. Lastly, resistance against oxidation and corrosion are relevant, most importantly when potential cycles in electrolyte are to provide the functionalization.

So far, good mechanical behavior of nanoporous materials with macroscopic dimensions -- several mm or more -- appears to be limited to precious nanoporous metals. Deformability in nanoporous Au has been observed in small-scale \cite{Biener2005,Volkert2006,Biener2006,Briot2018,Kim2018} and macro-scale testing \cite{Burckert2017,Jin2009,Ye2016,Mameka2016}. Also studies on nanoporous Pt \cite{Jin2007} or Pd \cite{Shi2017} have demonstrated plastic deformation behavior during macroscopic mechanical testing. While these materials have a good stability against the environment, they are costly.

Since the above precious metals are electropositive, they can be conveniently dealloyed in aqueous solution. Macroporous bodies of more abundant and affordable metals are obtained by liquid metal dealloying. Macroporous titanium, niobium, or stainless steel have been demonstrated in this way \cite{Kato2013,KatoNiobium2015,McCueMRSBull2018,OkulovKato2018}. Yet, dealloying in liquid metal is experimentally more onerous than aqueous dealloying. Furthermore, liquid metal dealloying typically yields ligament sizes well above hundred nanometers, preventing the study of many of the size effects that are required for functionalization or for studies of small-scale plasticity. Here, we demonstrate a convenient process, based on aqueous dealloying, towards nanoporous base metals which provide a low-cost alternative.

As Cu is rather electropositive, it is a natural candidate as an affordable material for dealloying processes based on corrosion in aqueous media. Nanoporous Cu has been prepared by dealloying from various alloys compositions that include -- next to Cu -- Mn \cite{Hayes2006,Jin2011a,Tang2016}, Al \cite{Kong2014,Cheng2012} and Zn \cite{Cheng2012,Egle2017}, respectively. However, while Cu in clean air forms a thick oxide layer that passivates the surface \cite{Keil2007}, Cu and its oxides are readily dissolved in numerous electrolytes when oxygen is present \cite{Ives1962}. Ni, on the other hand, forms a stable passive layer in air and in many acidic and alkaline corrosive environments \cite{Lambers1996,Davies1964,Rosenberg1968}. Several studies report the fabrication of nanoporous Ni using Mn-Ni precursor alloys \cite{Hakamada2009,Hakamada2009a,Dan2012,Hakamada2014,Cheng2017,Qiu2014}. Yet, the preparation of mechanically resilient nanoporous Ni remains a challenge, as is demonstrated by the brittle behavior observed in several studies \cite{Cheng2017,Qiu2014,Bai2016}.

Our preparation strategy explores alloys of Cu and Ni as promising candidates for combining the comparative ease of dealloying and the malleability of Cu with Ni's potential for passivation. We present nanoporous Copper-Nickel ("npCN") as low cost, deformable and inert nanoporous metal. Copper-Nickel, also known as \textit{Chinese}- or \textit{German Silver}, has been used in coin production for almost two millennia \cite{Needham1974,Rosenberg1968}. Its field of application has been extended to modern marine applications due to its excellent corrosion resistance \cite{Syrett1976,Bendall1995}.

Alloys of Cu, Ni and Mn are miscible at elevated temperature and so can be homogenized as solid solutions \cite{Gokcen1991,Gokcen1993,Bazhenov2013}, a prerequisite for making uniform nanoporous structures by dealloying \cite{Erlebacher2001}. We prepare npCN by electrochemically dealloying Mn-Cu-Ni, so that the preparation relies on base metals only. NpCN takes the form of macroscopic monolithic bodies, free of dealloying-induced cracks. There microstructure exhibits a network of ligaments with diameters of 13\,nm. Thermal annealing allows for controlled coarsening of the ligament size up to 40\,nm. We show that npCN is strong and deformable in compression.

\section{Preparing nanoporous Copper-Nickel}

The precursor alloy $\rm Cu_{20}Ni_{10}Mn_{70}$ was based on Cu and Ni metal wires (99.98+\,\% metal-base purity) and electrolytic Mn granulate (99.99\,\%). Prior to alloying, superficial oxides of the Mn were removed using 1\,M oxalic acid. Ingots of the precursor alloy were obtained by melting in a cold-crucible induction furnace. Afterwards, the precursor alloy was annealed at $850^\circ\rm C$ for $\sim 12$\,h and quenched in water. From this master alloy cylindrical samples were turned to a diameter of around $1.1\pm 0.05$\,mm using a lathe and cut into cylinders with a length of $1.8\pm 0.1$\,mm.

Electrochemical dealloying used a three electrode setup, with the precursor alloy clamped and electrically connected with gold wire, a graphite rod as counter electrode and Ag/AgCl (3\,M KCl, +210\,mV vs. standard hydrogen) as reference electrode. An aqueous solution prepared with high purity water (18.0\,$\rm M\Omega \rm cm$), 1\,M KCl ($\ge 99.5\,\%$ purity) and 10\,mM HCl ($\ge 99\,\%$) served as the electrolyte; its pH value was measured as 1.8. Potentiostatic dealloying at $-620\,\rm mV$ vs. Ag/AgCl was stopped when the current dropped to $<10\,\rm \mu A$. The material was then polarized at $-520\,\rm mV$ until the current diminished to zero, which removed further Mn. In order to avoid cracks induced by capillary forces during drying, the wet nanoporous samples were then immersed in Ethanol for around 2\,h followed by n-Pentane for $\geq 12$\,h and finally dried in air. This adopts a protocol from Ref~\cite{Bellet1998}.  A fraction of the samples were annealed in purified Argon gas (Oxygen and \BPChem{H\_2O} content $< 1$ ppm) at 400$^\circ $C for 20, 30 and 45\,min. Scanning electron microscopy was used to determine the mean ligament size, i.e. the diameter of connecting elements between nodes.

Compression tests with a constant engineering strain rate of ${\rm 10^{-4}\,s^{-1}}$ were performed using an universal testing machine equipped with a 2D digital image correlation (DIC) system (DaVis 8.3.1, LaVision). A pre-load of 1\,N was applied at the beginning of each measurement. The DIC setup enables in situ full field mapping of the sample's surface deformation in the camera's field of view. Images of the test setup and camera details are shown in Ref~\cite{Luhrs2016} and in the supporting online material of Ref~\cite{Luhrs2017}, respectively. Image correlation used a pixel size of $6-7\,\rm{\mu m}$, a step size of 8 pixels and a subset size of 25 pixels. Prior to testing, an airbrush system was used for applying a speckle pattern on the sample's lateral surface. This we found to greatly increase contrast and accuracy of the DIC measurement. The paint creates speckles with sizes from approximately 18 to 56\,$\rm{\mu m}$. Electron microscopy confirmed that paint did not enter the sample interior.

\section{Structural characterization}

\begin{figure*}[htb] \centering
	\includegraphics[width=1\textwidth]{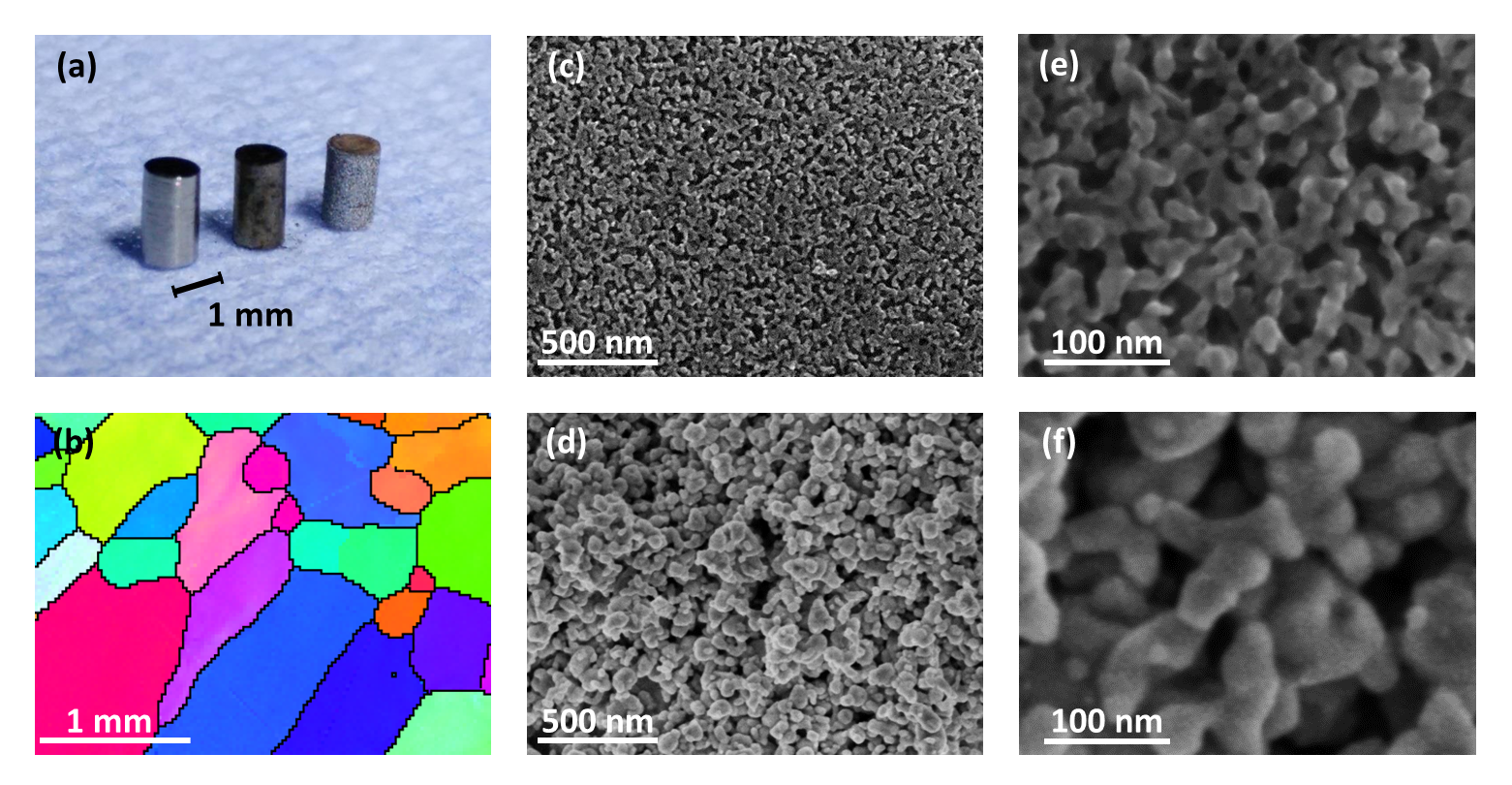} \protect
	\caption{ Structural characterization of nanoporous Copper-Nickel (npCN). \textbf{(a)} Photographs of (left to right) $\rm Cu_{20}Ni_{10}Mn_{70}$ precursor, as-dealloyed npCN and annealed npCN prepared for mechanical testing. Annealed sample has a speckled pattern applied; the sample was made from a slightly smaller master alloy than the remaining two. \textbf{(b)} Electron backscatter diffraction image (inverse pole figure coloring) Showing the microstructure of the master alloy. Grain boundaries are highlighted as black lines. \textbf{(c-f)} Scanning electron micrographs of cross-sectional areas in as-dealloyed npCN (c,\,e) and npCN annealed for 45\,min in Ar gas at $400^\circ$\,C ($L\approx 40$\,nm) (d,\,f) at different magnifications. Ligament sizes are 13 and 40 nm for as-dealloyed and annealed states, respectively.
	}
	\label{fig:REM}
\end{figure*}

The photographs of Fig~\ref{fig:REM} (a) illustrate geometry and optical appearance of samples in various stages of preparation: From left to right, a machined master alloy sample, an as-dealloyed npCN sample and an annealed npCN sample prepared for mechanical testing by spray-painting. As-dealloyed npCN takes on a matt dark brown color. Annealing brings the color closer to metallic copper, as can be seen on the paint-free end of the sample. The optical macrographs show monolithic bodies with no apparent macroscopic cracks. The annealed sample in Fig~\ref{fig:REM} (a) has an airbrush paint pattern for DIC applied to its side surfaces, as explained above.

Figure \ref{fig:REM} (b) illustrates the grain structure of the master alloy by means of an electron backscatter diffraction (EBSD) image.  Grain boundaries are highlighted. The analysis finds a single-phase face-centered cubic crystal structure, confirming successful homogenization  and quenching. At an average size of roughly 500\,$\rm{\mu m}$, the grains are quite large. The figure indicates texture in the grain shape which --  in view of the sample orientation in the ingot -- is consistent with the direction of solidification, from the walls of the cold crucible into the melt.

Figure \ref{fig:REM} shows scanning electron micrographs of as-dealloyed npCN (subfigures (c, e)) and npCN annealed for 45\,min (d, f) at different magnifications. The images were obtained on cross-sections of samples that were intentionally cleaved using a scalpel. It can be seen that the material exhibits a homogeneous bicontinuous structure, quite similar to that of macroscopic nanoporous gold \cite{Jin2009, Jin2011, Stenner2016}. Except for elongated pores, a few tens of nm in size, and in contrast to other studies on macroscopic nanoporous base metals \cite{Hayes2006, Cheng2017}, no dealloying-induced cracking is observed in npCN. In the as-dealloyed state the ligament size was identified as $13\pm 4$\,nm; measurement uncertainties represent the standard deviation. In analogy to nanoporous gold \cite{Li1992}, the ligament size  of npCN can be easily controlled by thermal coarsening. After annealing at 400$^\circ $C for 20, 30 and 45\,min, corresponding ligament sizes ensued to $22\pm 6$, $32\pm 8$ and $40\pm 14$\,nm, respectively.

Figures~\ref{fig:REM} (e) and (f) reveal that the ligaments of npCN are built of polyhedral elements; an indication of grooves can be seen where these elements are joined. The observation suggests that the ligament network is made of nanocrystallytes with a grain size that agrees with the ligament size. In this respect, npCN is distinguished from nanoporous gold, where the grain size is orders of magnitude larger than the ligament size \cite{Jin2009}. Grain boundary grooves might be seen as sites promoting pinch-off and loss of connectivity during coarsening. Yet, the ligament network of npCN appears to retain its connectivity and to coarsen in an essentially self-similar fashion. This is significant in the interest of mechanical performance of annealed samples with larger ligament size.

Density and solid volume fraction were determined through the assessment of the apparent sample volume by means of a measurement microscope and mass. Irrespective of the thermal treatment, density and solid volume fraction were measured to $2.75\pm 0.11\,\rm {g/cm^3}$ and $0.31\pm 0.01$, respectively. In opposition to nanoporous gold \cite{Parida2006,Luhrs2016,Luhrs2017} and macroscopic nanoporous nickel \cite{Cheng2017} npCN exhibits little shrinkage during dealloying and -- even more significantly -- during annealing.

Using energy dispersive X-ray analysis we find the Cu, Ni and Mn composition of dealloyed npCN to be $60\pm 6$\,at.\%, $32\pm 4$\,at.\% and $8\pm 2$\,at.\%, respectively.

\section{Mechanical properties}

\begin{figure}[htb] \centering
	\includegraphics[width=1\columnwidth]{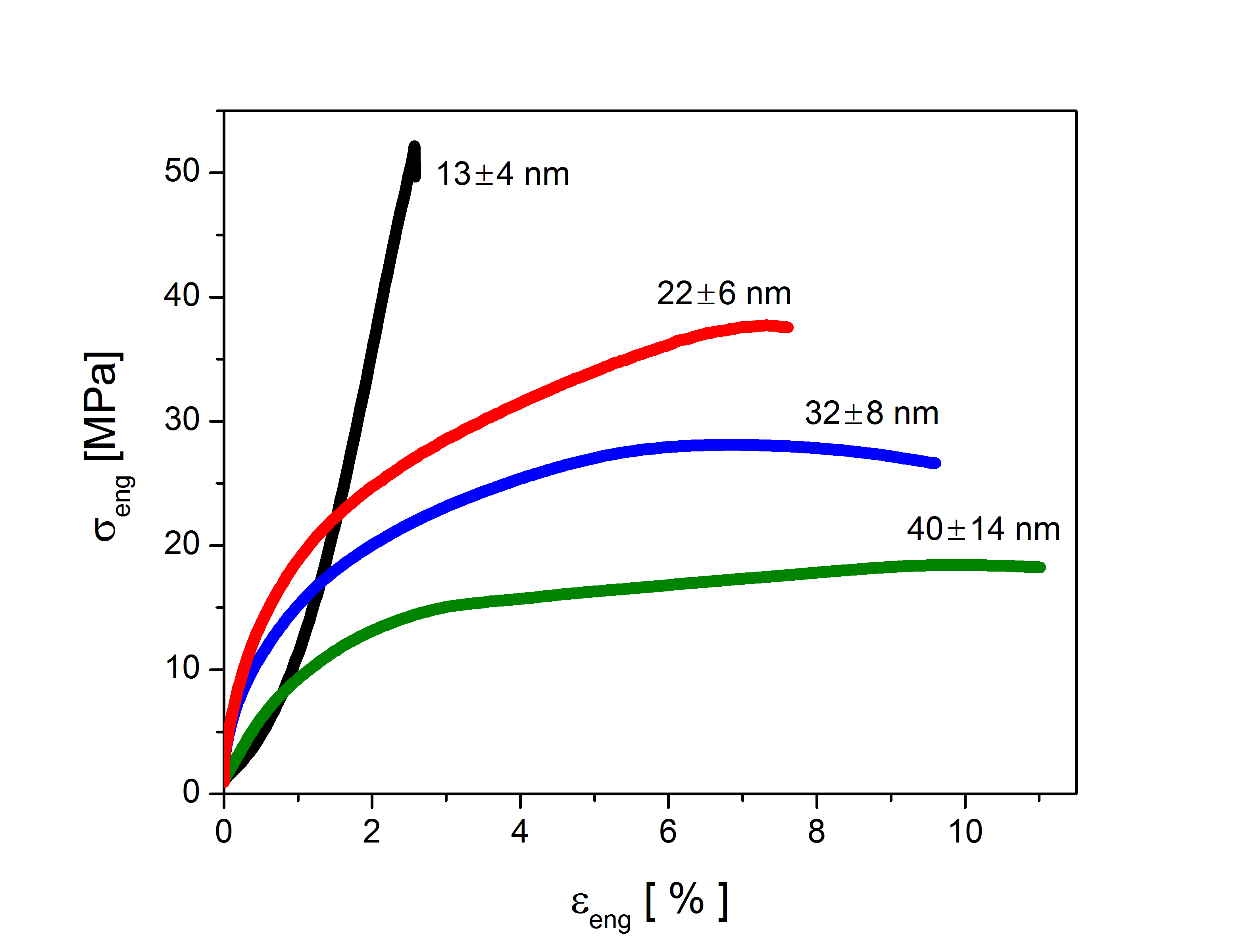} \protect
	\caption{ Compression tests on nanoporous Copper-Nickel (npCN). Engineering stress plotted vs. engineering strain for npCN with varying ligament sizes, \textit{L}. While the as-dealloyed material (black line) exhibits brittle deformation behavior, annealed npCN reveals significant ductility that increases with further coarsening of the ligament structure.
	}
	\label{fig:mech}
\end{figure}

Figure \ref{fig:mech} shows the results of continuous compression tests of npCN. The graph shows engineering stress versus engineering strain plots of as-dealloyed npCN (black line) as well as samples annealed for 20, 30 and 45\,min at 400$^\circ $C with corresponding ligament sizes. Analog to previous studies on nanoporous base metals \cite{Hayes2006,Kong2014,Bai2016,Cheng2017,Qiu2014,Chen2016}, as-dealloyed npCN shows brittle deformation behavior upon compression. Still, with a compressive and fracture stress of more than 50\,MPa, respectively, as-dealloyed npCN is strong compared to other nanoporous metals, such as those of Refs~ \cite{Balk2009,Jin2009,Burckert2017,Cheng2012,Chen2016}.

NpCN is also stronger than macroporous copper and nickel foams with a similar solid fraction, which exhibit yield stress values of around 5\,MPa for Ni- \cite{Yamada2007,Taylor2017} and up to 7\,MPa for Cu foams \cite{Hakamada2007,Parvanian2013}.

At the origin of the exceptional mechanical performance of our nanoporous base metal is on the one hand the absence of the dealloying-induced cracks that are observed in other studies \cite{Hayes2006,Cheng2017}. On the other hand, the high strength may also be a result of the small ligament size.  Studies on nanoporous gold  \cite{Biener2006,Jin2009,Burckert2017,Volkert2006} have confirmed the trend of "smaller is stronger" that originally emerged from microcolumn compression experiments on bulk gold  \cite{Greer2005,Greer2005a}.

In addition, due to the high surface area to volume ratio of npCN the nature of the surface may influence its mechanical performance. As described earlier, Cu and Ni readily form an oxide layer on the materials surface. During deformation, the dislocation motion can be impeded through pinning of dislocation endpoints at the adsorbed surface layer, a mechanism known as "adsorption locking" \cite{Westwood1962}. This effect has been exploited for tuning the flow stress of nanoporous gold. In such experiments, electrochemically controlled oxidation affords high strengthening through the reversible formation of an adsorbed oxide layer \cite{Jin2011}.

Another conceivable origin of the high strength of npCN might be solid solution strengthening. Indeed, macroscale compression tests on elemental nanoporous Cu (no solid solution hardening) report fracture- and yield strength not exceeding 6\,MPa \cite{Kong2014,Chen2016}, substantially less than what is achieved here with npCN. Microscale tests in Ref \cite{Cheng2012} find hardness values of around 20\,MPa for nanoporous Cu with a similar volume fraction. Hardness values of nanoporous metals have been identified with the yield strength \cite{Biener2006}, yet there are also arguments towards a more classic picture where the yield strength corresponds to 1/3 of the hardness \cite{Jin2009}. Furthermore, yield strengths obtained from microscale testing in nanoporous metals systematically exceed values found by macroscopic testing schemes \cite{Mameka2016}. This complicates comparisons between both testing methods. In any case, our macroscopic tests find the strength of npCN significantly higher than that of pure nanoporous Cu.

Compression tests of annealed npCN show considerable ductility that becomes more pronounced with increasing ligament size, $L$. With the notion of a "smaller is stronger" relation in mind the origin of the decrease in strength becomes obvious as the strengthening size effect diminishes with increasing $L$. Yet, even at $L \approx 40$\,nm and with a yield strength of $\ge$10\,MPa npCN exhibits a higher strength than macroporous Cu- and Ni-foams with a similar solid fraction \cite{Taylor2017,Yamada2007,Parvanian2013,Hakamada2007}. While high deformability in compression has been reported for some noble nanoporous metals \cite{Shi2017,Jin2007,Jin2009}, we here provide the first report of a  deformable nanoporous base metal. This is of importance since deformability -- and, with it, damage tolerance -- is a prerequisite for application.

\begin{figure*}[htb] \centering
	\includegraphics[width=0.7\textwidth]{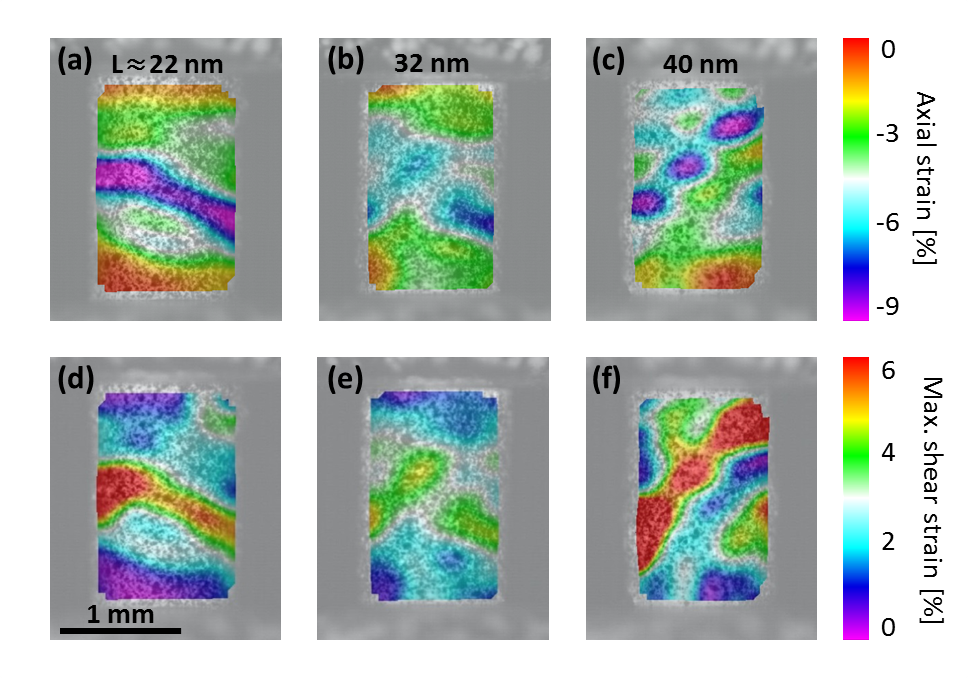} \protect
	\caption{ Strain maps of annealed nanoporous Copper-Nickel (npCN) under compressive loading at an average axial engineering strain of $4.4\,\%$. Axial-  \textbf{(a-c)} and maximum shear strain distribution  \textbf{(d-f)} shown for npCN with a ligament size of 22\,nm, 32\,nm and 40\,nm, respectively. Localized deformation clearly visible.
	}
	\label{fig:deformation}
\end{figure*}

Strain maps calculated with digital image correlation of annealed npCN during plastic deformation are shown in Figure \ref{fig:deformation}. Colors indicate the axial strain (top row of the figure) or the maximum two-dimensional shear strain on the sample surface. The images show npCN with $L \approx 22$\,nm, 32\,nm and 40\,nm, respectively, at an average compressive engineering strain of $4.4\,\%$ in loading direction. Axial strain maps (a-c), i.e. deformation in loading direction, find patches or bands of localized deformation with significantly higher straining compared to the surrounding material. An apparently similar deformation behavior is common in macroporous metallic foams in the form of "crush bands" \cite{Gibson1997,Bastawros2000}. However, the formation of crush bands in foams is carried by a collective collapse of adjacent cells, and it results in a stress drop. Compression tests of npCN (Fig.\,\ref{fig:mech}) reveal no stress maximum at early loading stages but on the contrary show distinct hardening during plastic deformation analog to macroscopic nanoporous gold \cite{Jin2009,Luhrs2016,Jin2011,Mameka2014,Luhrs2017,Mameka2016}. Also EBSD analysis of nanoporous Au at different loading stages shows no formation of localized crush bands \cite{Jin2009}. Instead inhomogeneous maximum shear strain distributions (d-f) suggest that the strain concentration during plastic deformation of annealed npCN can be attributed to shear deformation. This deformation ultimately leads to failure.

\section{Conclusion}

We have developed nanoporous Copper-Nickel (npCN), a new macroscopic nanoporous metal solely created from base metals. Single-phase $\rm Cu_{20}Ni_{10}Mn_{70}$ precursors were electrochemically dealloyed in  1\,M KCl and 10\,mM HCl electrolyte and carefully dried. Annealing the samples allowed for a controlled coarsening of the porous structures. Using electron microscopy we find homogeneous bicontinous network structures free of dealloying induced cracks with ligament sizes in the range of $13-40$\,nm, depending on the annealing conditions.

The mechanical performance of npCN was characterized via continuous compression testing. We find a high strength that exceeds values reported for macroporous metallic Cu- and Ni-foams, as well as pure nanoporous Cu with a similar volume fraction. We discussed the underlying strengthening mechanisms.

Most importantly we observe significant plastic deformation during mechanical testing, an essential advance for a nanoporous base metal produced by electrochemical dealloying. The deformability becomes more pronounced with increasing ligament size. We argue that both, the deformability as well as the low priced raw materials of nanoporous Copper-Nickel represent a important step for future industrial usage of tailor-made nanoporous metals.

\section*{Acknowledgment}

This work was funded by the Helmholtz-Gemeinschaft through the Helmholtz-CAS Joint Research Groups, project HCJRG-315.

\section*{References}

\bibliographystyle{model3-num-names}
\bibliography{bib}

\end{document}